\documentclass[%
 reprint,
onecolumn,
superscriptaddress,
 amsmath,amssymb,
 aps,
 pra,
floatfix,
]{revtex4-1}

\usepackage{newtxtext, newtxmath}
\usepackage[usenames, dvipsnames]{xcolor}
\usepackage{graphicx}
\usepackage{dcolumn} 
\usepackage{xpatch}
    \makeatletter \xpretocmd \start@align{\linenomathWithnumbers}{}{\fail}
\usepackage{mathtools, url}
\usepackage{multirow}
\usepackage{CJKutf8}
\usepackage[toc]{appendix}
\usepackage[ruled, linesnumbered]{algorithm2e}
\usepackage[unicode=true, bookmarks=true, bookmarksnumbered=false, bookmarksopen=false, breaklinks=false, pdfborder={0 0 1}, backref=false, colorlinks=false, hidelinks]{hyperref}
    \hypersetup{linkcolor=black, urlcolor=black, citecolor=black, pdfstartview={FitH}}

\usepackage{xr}
\usepackage{calc}
\usepackage{hyperref}
\usepackage{verbatim}
\usepackage{makecell}
\usepackage[ruled]{algorithm2e}
\usepackage{multirow}
\definecolor{darkgreen}{rgb}{0.0, 0.5, 0.0}
\usepackage{tikz}
\usetikzlibrary{quantikz2}
\usepackage{amssymb}
\usepackage{empheq}
\usepackage{float}

\usepackage[math]{cellspace}
\DeclareSymbolFont{CMlargesymbols}{OMX}{cmex}{m}{n}
\let\sumop\relax\let\prodop\relax
\DeclareMathSymbol{\sumop}{\mathop}{CMlargesymbols}{"50}
\DeclareMathSymbol{\prodop}{\mathop}{CMlargesymbols}{"51}

\SetSymbolFont{operators}{normal}{OT1}{ntxtlf}{m}{n}
\SetSymbolFont{operators}{bold}{OT1}{ntxtlf}{b}{n}

\newcommand{\ee}{\mathrm{e}}
\newcommand{\ii}{\mathrm{i}}

\renewcommand{\ket}[1]{| #1 \rangle}
\renewcommand{\bra}[1]{\langle #1 |}

\usepackage{booktabs}
\usepackage{multirow}
\usepackage{tabularx}
\usepackage{array}

\allowdisplaybreaks

\begin{document}
\title{Quantum computing for plasma physics in magnetic-confinement fusion}
\author{Chenjia Zhu} 
 \affiliation{State Key Laboratory for Turbulence and Complex Systems, School of Mechanics and Engineering Science, Peking University, Beijing 100871, China}%
 \affiliation{HEDPS-CAPT, Peking University, Beijing 100871, China}
\author{Shuai Xu} 
 \affiliation{Forschungszentrum J\"ulich GmbH, Institute of Fusion Energy and Nuclear Waste Management—Plasma
Physics, Partner of the Trilateral Euregio Cluster (TEC), 52425 J\"ulich, Germany}
\author{Yousheng Zhang} 
 \affiliation{Institute of Applied Physics and Computational Mathematics, Beijing, 100094, China}
 \affiliation{HEDPS-CAPT, Peking University, Beijing 100871, China}
 \affiliation{National Key Laboratory of Computational Physics, Beijing, 100088, China}
 \author{Yunfeng Liang} 
 \affiliation{Forschungszentrum J\"ulich GmbH, Institute of Fusion Energy and Nuclear Waste Management—Plasma
Physics, Partner of the Trilateral Euregio Cluster (TEC), 52425 J\"ulich, Germany}
\author{Yue Yang} 
 \email{yyg@pku.edu.cn}
 \affiliation{State Key Laboratory for Turbulence and Complex Systems, School of Mechanics and Engineering Science, Peking University, Beijing 100871, China}%
 \affiliation{HEDPS-CAPT, Peking University, Beijing 100871, China}


\begin{abstract}
Plasma simulations for magnetic-confinement fusion (MCF) involve high-dimensional kinetic dynamics, strong nonlinear coupling, complex geometries, and multiscale spatiotemporal behavior, imposing severe computational bottlenecks.
While quantum computing offers a potential alternative for accelerating plasma simulation problems, its practical advantage for fusion research has not yet been demonstrated. 
We review existing quantum computing studies across kinetic dynamics, wave interactions, equilibrium, and macroscopic stability, evaluating their hardware implementations and operational bottlenecks. 
Current studies remain largely restricted to linearized equations, reduced models, or isolated computational subroutines, leaving a substantial gap from realistic fusion simulations. 
We summarize key problem characteristics, including high dimensionality, structured dynamics, and measurement requirements, under which plasma applications are most likely to yield a quantum advantage. 
Guided by these considerations, we highlight potential approaches and present a roadmap toward practical quantum advantage of simulating plasma in MCF. 
\end{abstract} 

\maketitle

\section{Introduction}
\label{sec:intro}
\subsection{Plasma simulation in magnetic-confinement fusion}
\label{sec:introProblems}
\begin{figure*}
    \centering
    \includegraphics[width=\linewidth]{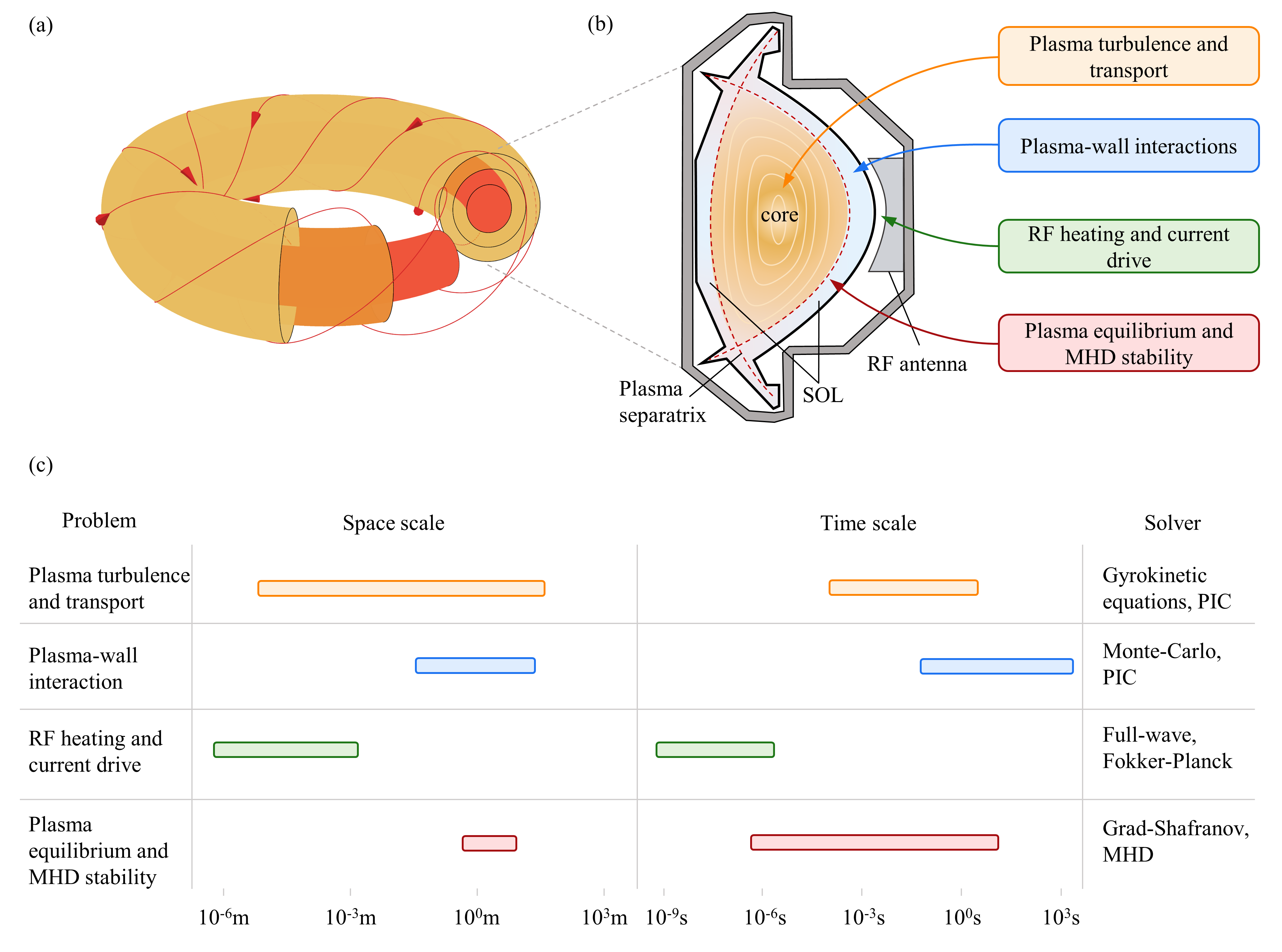}
    \caption{An overview of MCF and its computational challenges.
            (a) Schematic of MCF plasmas in toroidal plasma equilibrium. 
            Red curves denote representative magnetic field lines, while the translucent layers represent nested magnetic-flux surfaces.
            (b) Simplified poloidal cross-section of the SPARC tokamak adapted from Ref.~\cite{creely2020overview}, highlighting the plasma and vacuum vessel.
            Computational problems are associated with their respective regions.
            (c) Computational problems in MCF, and their representative spatial and temporal scales and commonly used numerical approaches~\cite{garcia2025challenges}. }
    \label{fig:plasmaOverview} 
\end{figure*}

Fusion energy offers a potential source of abundant and clean energy, but its realization requires the controlled confinement of plasmas at sufficiently high temperatures and for sufficiently long times to sustain fusion reactions. 
Among the major approaches to controlled fusion, magnetic-confinement fusion (MCF) uses strong magnetic fields to confine high-temperature plasmas in devices such as tokamaks, as illustrated in Fig.~\ref{fig:plasmaOverview}(a), and stellarators.

Magnetically confined fusion plasmas are complex, nonlinear systems spanning a wide range of spatial and temporal scales~\cite{fasoli2016computational}.
Computational modeling in MCF spans three broad areas: plasma dynamics, materials, and fusion engineering~\cite{garcia2025challenges}.
At the plasma level, the relevant scales range from electron-orbit motion, with characteristic spatial and temporal scales of approximately $10^{-5}$ m and $10^{-11}$ s, to macroscopic processes involving connection lengths of order $10^{2}$ m and current-diffusion times of order $10^{2}$ s.
When material and engineering processes are also considered, this range broadens further, spanning approximately $10^{-12}$--$10^{3}$ m in space and from picoseconds to years in time~\cite{garcia2025challenges}.
Moreover, these processes are often strongly coupled, making accurate prediction of fusion plasma evolution inherently a multiscale and multiphysics problem. 
Computational modeling and simulation therefore play an essential role in understanding plasma behavior, interpreting experiments, and supporting the design and operation of future fusion devices~\cite{campbell2025introduction}.

The multiscale and multiphysics nature of MCF gives rise to a broad hierarchy of computational problems.~\cite{garcia2025challenges}. 
Figure~\ref{fig:plasmaOverview} provides an overview of the device geometry, the associated plasma regions, and the representative computational problems and scales.
One of the most prominent is plasma turbulence and the associated transport processes.
Gradients in plasma temperature and density can drive instabilities at scales much smaller than the device, which develop into turbulence and enhance the radial transport of particles, heat, and momentum, thereby strongly affecting plasma confinement~\cite{garbet2010gyrokinetic, krommes2012gyrokinetic}.
Because fusion plasmas are strongly magnetized and weakly collisional, these processes are commonly described by gyrokinetic models derived from the Vlasov and Fokker-Planck equations and coupled to electromagnetic field equations~\cite{garbet2010gyrokinetic}. 
Current simulations typically solve gyrokinetic equations using a variety of numerical methods. 
Continuum codes employ spectral and finite-difference discretization, as implemented in GENE~\cite{jenko2000electron, goerler2011global}, while semi-Lagrangian schemes are used in codes such as GYSELA~\cite{grandgirard20165d}. 
Particle-based gyrokinetic simulations based on the particle-in-cell (PIC) method are also widely used, as exemplified by ORB5~\cite{lanti2020orb5}.

Another area of interest is plasma-wall interactions near the plasma boundary. 
In edge regions and scrape-off layer (SOL), the plasma is relatively cool and contains multiple species, including neutral particles and impurities released from plasma-facing materials, while also being influenced by both local and nonlocal turbulent transport~\cite{krieger2025scrape, schmid2020integrated}.
These processes are critical to heat and particle exhaust, component lifetime, and plasma purity~\cite{schmid2020integrated}.
Their numerical simulation typically requires a combination of complementary approaches. 
Edge-plasma transport is commonly modeled using coupled fluid and Monte-Carlo approaches, as implemented in SOLPS-ITER~\cite{wiesen2015new} and EMC3-EIRENE codes~\cite{feng19993d, reiter2005eirene}. 
Closer to material surfaces, kinetic and sheath-scale dynamics are treated using PIC methods, as implemented in the BIT1 code~\cite{tskhakaya2007particle, taccogna2008plasma}. 
At the material interface, Monte-Carlo codes such as ERO2.0~\cite{romazanov2020first} are further used to describe erosion, impurity transport, and redeposition.

Plasma equilibrium and magnetohydrodynamics (MHD) stability are also of great interest in overall fusion device design.
Therefore, it is essential to predict the overall plasma equilibrium~\cite{bandyopadhyay2025mhd}.
These processes occur at the device scale, and global MHD models and related Grad-Shafranov equations are usually used in numerical simulations~\cite{takeda1991computation}.
The equilibrium problem of a given MHD model is typically transformed into a minimization problem using finite-difference or spectral methods and then solved numerically~\cite{takeda1991computation, helander2014theory}.
An example code for ideal 3D MHD equilibrium modeling is DESC~\cite{dudt2020desc}, while SPEC~\cite{hudson2012computation} is a notable code for solving multi-region relaxed MHD models.

Apart from these, radio-frequency (RF) heating and current drive are also of great interest. 
In fusion devices, electromagnetic waves are widely used for plasma heating and current drive, since Ohmic heating becomes less effective as the plasma temperature increases~\cite{cairns1991radiofrequency, heuraux2015simulation}.
The wave energy and momentum are transferred to plasma particles through wave-particle interactions and subsequently redistributed through collisions. 
RF wave propagation and absorption can be simulated with full-wave solvers such as AORSA~\cite{jaeger2001all}. 
Fokker-Planck solvers such as CQL3D are used to solve the collisional and quasi-linear evolution of distribution functions of ions and electrons, and calculate the resulting driven current~\cite{harvey1992cql3d}.
In addition, RF heating involving strong nonlinearities and edge effects~\cite{heuraux2015simulation, hillairet2023review} is modeled using nonlinear full-wave and sheath models, as implemented in codes such as Petra-M~\cite{bertelli2023benchmark}.

This review primarily focuses on MCF plasmas, which are characterized by high temperatures, strong magnetization, weak collisionality, and pronounced multiscale coupling. 
In contrast, inertial confinement fusion is predominantly modeled at the macroscopic scale using radiation-hydrodynamic fluid equations, with major computational challenges arising from strong compression, shock propagation, radiation transport, and their multiphysics coupling~\cite{zhou2025instabilities, betti2016inertial}. 
Since this review focuses primarily on computational problems associated with plasma dynamics, inertial confinement fusion is not discussed in detail.
Despite this focus, many of the governing equations and computational structures are shared with other plasma regimes, including high-energy-density~\cite{tzeferacos2015flash, lezhnin2025particle} and low-temperature plasmas~\cite{kim2005particle}, dusty plasmas~\cite{donko2019molecular}, and plasmas encountered in space~\cite{toth2012adaptive} and astrophysical~\cite{palmroth2025vlasov} environments.
The models and algorithms reviewed here, including methods for kinetic equations, electromagnetic wave propagation, and magnetohydrodynamic models, are therefore not limited to magnetic-confinement fusion and can potentially be extended to these related settings. 
Such extensions generally require adapting the geometry, boundary and initial conditions, characteristic physical parameters, and observables to the specific plasma regime.

\subsection{Major computational challenges in plasma physics}
\label{sec:introChallenges}
Plasma simulations in MCF are computationally demanding because they combine broad spatial and temporal scale separations with nonlinear, multiphysics, and geometry-dependent numerical challenges.~\cite{fasoli2016computational}. 
These include high-dimensional kinetic dynamics, nonlinear and self-consistent coupling, complex geometries and boundary conditions, and the coupling of processes occurring on widely separated scales~\cite{garcia2025challenges}. 
The dominant computational bottleneck therefore varies substantially among different problems in plasma physics. 

First, plasma turbulence and transport is one of the most computationally demanding parts in fusion research~\cite{garcia2025challenges, apollonatos2012helios}. 
A major challenge stems from the high dimensionality of gyrokinetic models, which imposes substantial computational and memory costs particularly when multiple particle species and electromagnetic effects are included.
Nonlinear processes further increase computational costs through the self-consistent coupling between particle distributions and electromagnetic fields, along with nonlinear collisional effects in plasma transport.
In addition, microscopic turbulent structures interact with slowly evolving macroscopic plasma profiles to generate complex coupling across widely separated spatial and temporal scales~\cite{abel2013multiscale}. 
Therefore, predictive transport simulations must capture this self-consistent coupling over long time scales, making global simulations challenging due to high computational costs and accumulated numerical noise~\cite{parker2018bringing}.

Second, plasma-wall interactions pose computational challenges involving complex edge dynamics, geometries, and model coupling.
Turbulent fluctuations in these regions become comparable to equilibrium quantities, which complicates the standard separation between slowly varying background profiles and small fluctuations and requires non-perturbative treatments that retains the large fluctuations and strong gradients~\cite{beadle2020understanding, schwander2024global}.
In addition, plasma density, temperature, and other properties can vary strongly across the edge region, giving rise to pronounced nonlinearities, steep gradients, and numerical stiffness. 
Realistic simulations must also account for complex magnetic and material geometries, which further complicate spatial discretization and boundary conditions~\cite{schwander2024global}. 
Finally, plasma-wall interactions introduce the coupling of distinct physical models such as impurities and erosion~\cite{wiesen2017plasma}, which further increases computational cost and complexity.

Third, plasma equilibrium under realistic geometries and physical models continues to pose computational challenges.
MHD equilibrium incorporates the balance among plasma pressure, magnetic fields, and current, which involves nonlinear and self-consistent coupling between different fields~\cite{bandyopadhyay2025mhd}. 
Free-boundary equilibrium introduces additional complexity, as the plasma boundary must be determined self-consistently with the external magnetic fields~\cite{conlin2024high}. 
Moreover, the computational cost further increases in 3D geometries because of the increased number of degrees of freedom and the loss of axisymmetry~\cite{hindenlang2025computing}. 
Equilibrium reconstruction is also frequently performed during plasma discharges to provide information for monitoring and control~\cite{ferron1998real}. 
Although real-time numerical applications have already been developed for axisymmetric tokamak equilibrium~\cite{ferron1998real}, further acceleration is essential to extend real-time calculation to higher-fidelity models and more complex geometries.

Fourth, for RF heating and current drive, the main computational challenges arise from the propagation and absorption of electromagnetic waves in strongly inhomogeneous and anisotropic magnetized plasmas~\cite{heuraux2015simulation}. 
High-fidelity simulations often require full-wave calculations with realistic geometries, while the deposited wave energy and driven current depend on wave-particle interactions and the evolution of particle distributions~\cite{shiraiwa2026full}. 
Coupling electromagnetic-wave solvers with kinetic models therefore introduces additional computational cost, particularly when nonlinear plasma responses and complex boundary effects must be accounted for~\cite{frank2022verifying}.

The multiscale, nonlinear, and multiphysics difficulties described above have motivated the development of a hierarchy of plasma models, which include particle, kinetic, and gyrokinetic descriptions to fluid, MHD, and reduced models. 
Recent advances in high-performance computing, parallel computing, and accelerator hardware such as GPUs have greatly accelerated research in numerical simulations of plasma problems~\cite{wright2024developing}. 
Nevertheless, further advances in high-performance computing still involve tradeoffs among portability, scalability, and I/O optimization~\cite{garcia2025challenges}. 
In summary, high-fidelity simulations of many experimentally relevant plasma phenomena remain computationally expensive across nearly all areas of plasma physics, and there remains a strong demand for computational tools capable of supporting practical fusion-device applications.

\subsection{Scope and objectives}
The computational challenges discussed above motivate the exploration of alternative computing paradigms, among which quantum computing provides a potentially promising route for accelerating certain plasma problems. 
Quantum computing for plasma physics is still a relatively new research direction. 
Several previous reviews have discussed the intersection of quantum computing and plasma physics~\cite{joseph2023, dodin2020applications, amaro2023living}, with much of their emphasis placed on introducing basic concepts of quantum computing and representative quantum algorithms relevant to plasma applications.

This review examines quantum computing from the perspective of computational tasks in magnetic-confinement fusion, with particular attention to the physical models, algorithmic approaches, and implementation status of existing studies. 
We first systematically review existing quantum algorithms developed for plasma-physics equations and applications, with attention to the physical models and numerical structures they target. 
We then identify plasma problems for which quantum computing provides meaningful computational advantages and examine the conditions under which such advantages could be realized. 
We also assess existing hardware demonstrations and available resource estimates to evaluate the current level of practical feasibility. 
Finally, based on the progress and limitations identified in existing studies, we discuss promising research directions and possible roadmaps toward physically relevant quantum plasma simulations.

The remainder of this review is organized as follows. Section~\ref{sec:qcIntro} introduces the basic concepts and algorithms of quantum computing relevant to plasma simulation. Section~\ref{sec:currentResearch} reviews existing progress in quantum computing for plasma physics. Section~\ref{sec:roadmaps} discusses the major challenges, promising directions, and possible routes for future development, followed by conclusions in Sec.~\ref{sec:Conclusions}.

\section{Introduction to quantum computing}
\label{sec:qcIntro}
\subsection{Quantum representation and computation}
Quantum computing provides a promising alternative computational paradigm for addressing complex problems in plasma physics.
By exploiting quantum phenomena such as superposition, entanglement, and interference, quantum algorithms offer potential computational advantages over their classical counterparts~\cite{Nielsen2010Quantum}.
For example, quantum states can compactly represent high-dimensional data, which is attractive for kinetic plasma simulations involving large phase spaces.
A quantum workflow typically consists of initial state preparation, quantum evolution, and final state measurement.
We briefly introduce the basic concepts of quantum computing, quantum algorithms relevant to plasma physics, and the major challenges toward practical quantum advantage.

A quantum computation begins by encoding input data into an initial quantum state.
A single-qubit state is represented by $\ket{\psi} = \alpha \ket{0} + \beta \ket{1} = (\alpha, \beta)^\mathrm{T}$, 
where $\ket{0}$ and $\ket{1}$ are computational basis states, $\alpha$ and $\beta$ are probability amplitudes, and $|\alpha|^2 + |\beta|^2 = 1$.
The normalization condition ensures unit total measurement probability.
Multiple-qubit systems are described in tensor-product Hilbert spaces. 
An $n$-qubit state is expressed as $\ket{\psi} = \sum_{j = 0}^{2^n - 1} \alpha_j \ket{j}$,
where $\ket{j}$ denotes an $n$-qubit computational basis state, and $\sum_{j = 0}^{2^n - 1} |\alpha_j|^2 = 1$.
Thus, $n$ qubits span a $2^n$-dimensional Hilbert space, enabling high-dimensional data to be represented compactly.
In practical applications, classical data need to be encoded into quantum states using suitable encoding strategies depending on the structure of the problem.
The efficiency of this encoding and the associated state-preparation cost can significantly affect the overall computational advantage.

After data encoding, quantum operations are applied to the prepared state through a quantum circuit.
The evolution of a closed quantum system is governed by the linear Schr\"odinger equation $\ii \hbar \partial_t \ket{\psi} = H\ket{\psi}$, where $H$ is a Hermitian Hamiltonian.
For a time-independent $H$,
\begin{equation}
\ket{\psi'} = U\ket{\psi}, \quad U(t) = \ee^{-\ii Ht/\hbar},
\label{eq:linearSchor}
\end{equation}
where $U$ is unitary and satisfies $U^\dagger U=I$.
In the circuit model, such unitary transformations are implemented through quantum gates, with elementary single- and multi-qubit gates combined to form a quantum circuit.
Quantum algorithms also frequently employ oracles, which are specially designed unitary operations for data access or operator implementation.
Apart from the circuit model, certain problems can also be implemented using the quantum systems whose native dynamics realize the desired evolution.

The final results of a quantum computation are extracted through quantum measurement.
For a quantum state $\ket{\psi}=\sum_j a_j\ket{j}$, measurement in the computational basis yields $\ket{j}$ with probability $|a_j|^2$.
Physical quantities are generally represented by Hermitian observables $O$, whose expectation values are $\langle O\rangle=\bra{\psi}O\ket{\psi}$.
Since a single measurement yields only one outcome, repeated measurements are generally required to estimate probabilities or expectation values accurately~\cite{Nielsen2010Quantum}. 
Therefore, extracting full classical information from the final state can be expensive, and quantum algorithms are more favorable when only a limited number of observables need to be measured.

Real quantum hardware is inevitably affected by noise, including decoherence, gate errors, and measurement errors.
Its computational capability is commonly characterized by qubit count, gate fidelity, coherence time, qubit connectivity, and executable circuit depth, which respectively describe the system size, operation accuracy, preservation of quantum information, available qubit interactions, and complexity of executable circuits.
Current noisy intermediate-scale quantum (NISQ) devices operate with limited circuit depth and without full error correction~\cite{preskill2018quantum}. 
For reliable large-scale computation, quantum error correction encodes logical qubits using multiple physical qubits to detect and correct errors~\cite{shor1995scheme}. 
Such fault-tolerant implementations introduce substantial overhead in physical qubits and operations~\cite{fowler2012surface}, and practical resource estimates should therefore distinguish between logical and physical resources.



\subsection{Quantum algorithms for linear problems}
As quantum-state evolution is governed by the linear Schr\"odinger equation, quantum computing provides a natural framework for problems with linear and structured mathematical formulations. 
Quantum algorithms have been developed for several important tasks, including Hamiltonian simulation, linear-system solving, linear differential equations, and eigenvalue estimation. 

There are several primitives across specific problems.
A basic primitive in quantum linear algorithms is block encoding.
To process non-unitary matrices on unitary quantum circuits, block encoding embeds a general matrix $A$ into a larger unitary matrix $U_A$ as a subblock, where $U_A$ satisfies
\begin{equation}
   (\langle 0| \otimes I) U_A (|0\rangle \otimes I) = A / \alpha,
\end{equation}
and $\alpha \geq \|A\|$ is the normalization factor. 
Another widely used primitive is the linear combination of unitaries (LCU), which represents a target operator as a weighted sum of unitary operators~\cite{childs2012hamiltonian}. 
Quantum signal processing (QSP) implements polynomial transformations associated with unitary operators, while quantum singular value transformation (QSVT) extends this framework to polynomial transformations of the singular values of block-encoded matrices~\cite{gilyen2019quantum, low2019hamiltonian}. 
Collectively, these primitives underpin the higher-level algorithms detailed below.

Building on these common primitives, Hamiltonian simulation is one of the fundamental tasks in quantum computing, aiming to implement the unitary time evolution $U(t)$ in Eq.~\eqref{eq:linearSchor}.
One of the most widely used approach are product-formula methods, such as Lie-Trotter formula~\cite{trotter1959product, lloyd1996universal}.
For a Hamiltonian decomposed as $H=\sum_l H_l$, product-formula methods approximate the full evolution by a sequence of simpler evolutions.
Another class of methods approximates $U(t)$ using truncated series and implements the resulting approximation through matrix polynomial techniques. 
These series can be implemented using LCU~\cite{berry2015simulating, kieferova2019simulating} and QSVT~\cite{gilyen2019quantum, low2019hamiltonian}.

Another important class of quantum algorithms is quantum linear system algorithms (QLSAs), which aim to provide potential speedups for solving linear systems $Ax=b$.
The most well-known QLSA is the Harrow-Hassidim-Lloyd (HHL) algorithm, which prepares a solution state $\ket{x}\propto A^{-1}\ket{b}$ with controlled rotations~\cite{harrow2009quantum}. 
Later methods implement matrix inversion through more general matrix-function techniques by approximating the function $f(x)=1/x$ using suitable series or polynomials, which are then implemented through LCU~\cite{childs2017quantum} or QSVT~\cite{gilyen2019quantum}. 
QLSAs generally output the solution as a quantum state rather than an explicit classical vector, and their practical advantage also depends on additional costs such as state preparation and measurement.

To solve non-unitary linear differential equations, quantum algorithms have been developed to recast non-unitary dynamics into quantum-compatible formulations. 
One approach discretizes the original differential equation and reformulates it as a linear system, which is solved with QLSAs~\cite{berry2014high, berry2017quantum, childs2020quantum}.
Alternatively, non-unitary dynamics can be mapped to or represented by unitary evolutions that can be implemented through Hamiltonian simulation. 
Representative approaches include Schr\"odingerization~\cite{jin2023quantum, jin2024quantum, jin2025schrodingerization}, and linear combination of Hamiltonian simulations~\cite{an2023linear, an2026quantum, yang2025circuit, Meng2026_Toward}.
For specific problems, techniques such as quantum spin encoding~\cite{meng2023quantum, meng2024quantum, meng2024simulating, su2026quantum}, lattice gas cellular automata~\cite{yepez2001quantum, Wang2025_Quantum}, and qubit lattice algorithms (QLA)~\cite{yepez2002efficient} use specific quantum representations to construct quantum-compatible evolutions.

Quantum eigenvalue algorithms address spectral problems essential for stability and mode analyses, spanning both fault-tolerant and near-term hardware paradigms. 
In the fault-tolerant regime, quantum phase estimation extracts eigenvalues from phase accumulated under controlled unitary evolution~\cite{kitaev1995quantum}. 
Conversely, for near-term devices, the variational quantum eigensolver provides a hybrid alternative that approximates eigenstates by optimizing parameterized quantum states~\cite{peruzzo2014variational, Tilly2022variational}.

Several additional quantum primitives are also relevant to plasma applications. 
The quantum Fourier transform implements the discrete Fourier transform on quantum states and serves as a basic component of algorithms such as quantum phase estimation~\cite{coppersmith2002approximate}. 
Quantum amplitude estimation provides a quadratic speedup in estimating expectation values and probabilities over classical sampling~\cite{brassard2000quantum}. 
This advantage is particularly effective for quantum Monte-Carlo methods and aligns well with plasma problems that require extracting only a limited number of statistical quantities.

\subsection{Quantum algorithms for nonlinear problems}
Extending quantum algorithms from linear systems to nonlinear dynamics remains fundamentally difficult, as standard quantum circuits execute linear unitary operations that cannot naturally accommodate nonlinear evolution~\cite{tennie2025quantum}.
A central goal is therefore to reformulate nonlinear dynamics into quantum-compatible forms without introducing prohibitive computational overhead.

Existing approaches to bridge this gap generally fall into two categories. 
The first category reformulates nonlinear dynamics as linear evolution in an enlarged space.
Carleman linearization achieves this for polynomial systems via monomial expansions~\cite{liu2021efficient, itani2022analysis, sanavio2024three, costa2025further}, whereas the Koopman operator framework accommodates general nonlinearities through linear observable dynamics in function space~\cite{giannakis2022embedding, zhang2025data, gan2025provably}. 
Related homotopy methods similarly decompose equations into linear sequences~\cite{xue2021quantum, liao2024ham, xue2025quantum}.
More specialized approaches construct quantum representations of the original dynamics. 
For example, second quantization maps nonlinear classical dynamics to linear quantum evolution in an enlarged Fock space, from which the original nonlinear dynamics are recovered in the classical limit~\cite{may2024algebraic}.
While these approaches provide general pathways to convert nonlinear dynamics to linear dynamics, they generally require truncation of the enlarged representation, which introduces additional overhead and errors.
%

The second category relies on hybrid quantum-classical frameworks, which approximate nonlinear dynamics through classical feedback or variational optimizations.
Variational quantum algorithms for near-term devices use parameterized quantum circuits together with classical optimization~\cite{Cerezo2021variational, lubasch2020variational, sarma2024quantum}.
They have been applied to a variety of nonlinear dynamical problems and quantum computing procedures~\cite{wang2026simulating, jesus2026time, kocher2025numerical, su2026quantum, zhu2025quantum}. 
Alternatively, other hybrid strategies isolate quantum-suitable subproblems by decoupling linear and nonlinear dynamics through operator splitting~\cite{kocher2025numerical, weng2026quantum} or by exploiting problem-specific mathematical structures such as Ginzburg-Landau potentials~\cite{jin2026hybrid} and Lax-pair scattering~\cite{zhu2026quantum}.
Despite their flexibility, their practical application is constrained by repeated measurements, classical-feedback and optimization overheads, and the applicability of problem-specific formulations~\cite{Cerezo2021variational, lewis2024limitations}.

\subsection{Obstacles in quantum computing}
While the quantum algorithms discussed above provide promising approaches to a range of linear and nonlinear plasma-related problems, several obstacles remain before these theoretical advantages can translate into practical benefits for plasma simulations~\cite{tennie2025quantum}. 
Current NISQ devices operate with limited scale and without full fault-tolerant protection~\cite{preskill2018quantum}.

Three closely related challenges need to be addressed to achieve reliable large-scale quantum computing~\cite{vukvsic2026comparative}.
First, decoherence and operational errors limit circuit accuracy and depth, as environmental interactions and imperfections in initialization, gate operations, and readout make it difficult to preserve quantum information throughout a computation~\cite{preskill2018quantum,vukvsic2026comparative}.
Second, scaling quantum processors introduces increasing engineering complexity, as maintaining high operational fidelity across larger systems requires precise control, reliable connectivity, and stable calibration of many qubits~\cite{vukvsic2026comparative}.
Third, quantum error correction introduces substantial resource overhead, as encoding and protecting each logical qubit requires multiple physical qubits and repeated error-detection and correction cycles~\cite{shor1995scheme,fowler2012surface}.

Apart from hardware challenges, practical applications require efficient preparation of initial quantum states, data encoding, and extraction of useful results through measurement~\cite{preskill2018quantum}.
Algorithmic speedups also depend on problem-specific conditions, such as operator structure, conditioning, and required accuracy~\cite{harrow2009quantum,childs2017quantum}.
Practical quantum advantage therefore needs to be assessed over the complete computational workflow.



\section{Current quantum algorithms for plasma physics}
\label{sec:currentResearch}
We review existing quantum computing studies for major plasma physics problems, focusing on the physical models considered, the quantum approaches employed, and their current level of implementation.
These studies are summarized in Tab.~\ref{tab:curentAlgorithms}. 

\begin{table*}[t]
\centering
\caption{Summary of quantum-computing approaches for plasma physics. Superscripts denote the main algorithmic approaches: H denotes Hamiltonian simulation, L quantum linear system algorithms, D quantum solvers for linear differential equations, E linearization or embedding, V variational or quantum-classical hybrid approaches, S quantum spectral or eigenvalue algorithms, and R quantum representations. 
The symbol $\dagger$ indicates studies involving physical quantum hardware implementations.}
\label{tab:curentAlgorithms}

\renewcommand{\arraystretch}{1.2}

\newcommand{\qpProblemCell}[1]{%
    \parbox[c]{0.22\textwidth}{\raggedright #1}}
\newcommand{\qpModelCell}[1]{%
    \parbox[c]{0.27\textwidth}{\raggedright #1}}
\newcommand{\qpStudyCell}[1]{%
    \parbox[c]{\dimexpr0.51\textwidth-4\tabcolsep\relax}{\raggedright #1}}
\newcommand{\qpHeaderCell}[2]{%
    \parbox[c][1.8\baselineskip][c]{#1}{\raggedright\textbf{#2}}}
\newcommand{\qpRowGap}{\addlinespace[5pt]}
\newcommand{\qpRowGapSmall}{\addlinespace[2pt]}

\newcommand{\HS}{\textsuperscript{H}}
\newcommand{\QLSA}{\textsuperscript{L}}
\newcommand{\DE}{\textsuperscript{D}}
\newcommand{\EMB}{\textsuperscript{E}}
\newcommand{\HYB}{\textsuperscript{V}}
\newcommand{\REP}{\textsuperscript{R}}
\newcommand{\SPE}{\textsuperscript{S}}
\newcommand{\HRD}{\textsuperscript{$\dagger$}}

\begin{tabular}{@{}ll@{}}
\toprule
\qpHeaderCell{0.22\textwidth}{Problem}
&
\begin{tabular}[c]{@{}ll@{}}
    \qpHeaderCell{0.27\textwidth}{Model}
    &
    \qpHeaderCell{\dimexpr0.51\textwidth-4\tabcolsep\relax}{Quantum computing studies}
\end{tabular}
\\
\midrule

\qpProblemCell{Plasma turbulence and transport}
&
\begin{tabular}[c]{@{}ll@{}}
    \qpModelCell{Linear collisionless kinetic models}
    &
    \qpStudyCell{Engel et al.~\cite{engel2019}\HS, Ameri et al.~\cite{ameri2023}\HS, Toyoizumi et al.~\cite{toyoizumi2024hamiltonian}\HS, Paul et al.~\cite{paul2026efficient}\HS, Sugama et al.~\cite{sugama2026schr}\REP, Novikau et al.~\cite{novikau2024encoding}\QLSA, Goldfriend et al.~\cite{goldfriend2025implementation}\QLSA}
    \\
    \qpRowGap
    \qpModelCell{Linear collisional kinetic models}
    &
    \qpStudyCell{Ameri et al.~\cite{ameri2023}\DE, Amaro et al.~\cite{amaro2025variational}\HYB}
    \\
    \qpRowGap
    \qpModelCell{Nonlinear kinetic models}
    &
    \qpStudyCell{Vaszary et al.~\cite{vaszary2024}\EMB\QLSA, Berntson et al.~\cite{berntson2026end}\EMB, Christensen et al.~\cite{christensen2026improved}\EMB\DE, May \& Qin~\cite{may2025}\REP, Tronci \& Joseph~\cite{tronci2021koopman}\EMB, Higuchi et al.~\cite{higuchi2025}\HYB}
    \\
    \qpRowGapSmall
\end{tabular}
\\

\midrule

\qpProblemCell{RF heating and current drive}
&
\begin{tabular}[c]{@{}ll@{}}
    \qpModelCell{Linear wave propagation}
    &
    \qpStudyCell{Novikau et al.~\cite{novikau2022quantum}\HS,  Koukoutsis et al.~\cite{koukoutsis2023}\HS, Koukoutsis et al.~\cite{koukoutsis2025quantum}\HS\DE, Hizanidis et al.~\cite{hizanidis2025spacetime}\REP, Novikau et al.~\cite{novikau2023simulation}\QLSA, Sundar et al.~\cite{sundar2026simulating}\REP\HS\HRD}
    \\
    \qpRowGap
    \qpModelCell{Nonlinear wave interactions}
    &
    \qpStudyCell{Shi et al.~\cite{shi2021simulating}\REP\HS\HRD, May \& Qin~\cite{may2023quantum}\REP, May \& Qin~\cite{may2024nonlinear}\REP, Shi et al.~\cite{shi2024simulating}\REP\HS\HRD}
    \\
    \qpRowGap
    \qpModelCell{Chaotic particle dynamics}
    &
    \qpStudyCell{Porter \& Joseph~\cite{porter2025impact}\HS\HRD}
    \\
    \qpRowGapSmall
\end{tabular}
\\

\midrule

\qpProblemCell{Plasma equilibrium and MHD stability}
&
\begin{tabular}[c]{@{}ll@{}}
    \qpModelCell{MHD equilibrium and stability}
    &
    \qpStudyCell{Ameri et al.~\cite{ameri2023quantum, ameri2024mhd, ameri2025quantum}\SPE, Mastorakis et al.~\cite{mastorakis2026quantum}\HYB}
    \\
    \qpRowGap
    \qpModelCell{Reduced MHD / fluid dynamics}
    &
    \qpStudyCell{Vahala et al.~\cite{vahala2003lattice}\REP\HYB, Yepez et al.~\cite{yepez2005quantum}\REP\HYB, Higuchi et al.~\cite{higuchi2025quantum}\EMB\HS}
    \\
    \qpRowGapSmall
\end{tabular}
\\

\midrule

\qpProblemCell{Other applications}
&
\begin{tabular}[c]{@{}ll@{}}
    \qpModelCell{Quantum-assisted PIC}
    &
    \qpStudyCell{Hegde et al.~\cite{hegde2025hybrid}\HYB}
    \\
    \qpRowGap
    \qpModelCell{Plasma diagnostics}
    &
    \qpStudyCell{Wang et al.~\cite{wang2026validating}\HYB}
    \\
    \qpRowGapSmall
\end{tabular}
\\

\bottomrule
\end{tabular}

\end{table*}

\subsection{Plasma turbulence and transport}
Among the problems introduced in Sec.~\ref{sec:introProblems}, kinetic plasma dynamics has received the most attention in existing studies of quantum algorithms, because kinetic models are high-dimensional and often retain exploitable linear structures. 
In MCF plasmas, kinetic dynamics is commonly described by the Vlasov equation with collisions
\begin{equation}
    \frac{\partial f_s}{\partial t}
    + \mathbf{v} \cdot \nabla_{\mathbf{x}} f_s
    + \frac{q_s}{m_s}
    \left( \mathbf{E} + \mathbf{v} \times \mathbf{B} \right)
    \cdot \nabla_{\mathbf{v}} f_s
    = C[f_s],
    \label{eq:vlasov}
\end{equation}
where $f_s(\mathbf{x},\mathbf{v},t)$ is the distribution function of species $s$, 
$\mathbf{v}$ is the velocity-space coordinate, 
$q_s$ and $m_s$ are the particle charge and mass, respectively, $\mathbf{E}$ and $\mathbf{B}$ denote the electric and magnetic fields, and $C[f_s]$ represents collisional effects.
The second term describes particle transport in physical space, and the third term %
describes transport in velocity space driven by the Lorentz force. 
The collision operator $C[f_s]$ accounts for interactions between particles and takes different forms depending on the collision model adopted.
Equation~\eqref{eq:vlasov} is often solved with Maxwell's equations to obtain $\mathbf{B}$ and $\mathbf{E}$, where the essential charge and current densities are derived self-consistently from $f_s$. 
In electrostatic models, the electromagnetic coupling is often simplified to Gauss's law, or equivalently the Poisson equation for the electrostatic potential.

Early quantum algorithms for kinetic plasma dynamics mainly focused on the linearized Vlasov-Poisson equations.
For plasmas close to equilibrium, the dynamics can be described in terms of small perturbations to the equilibrium distribution.
Under the electrostatic approximation, magnetic field perturbations are neglected and the background equilibrium fields $\mathbf{E}_0 $ and $\mathbf{B}_0$ are considered to be 0.
The electric field is then determined self-consistently from the perturbed charge density.
The resulting linearized Vlasov-Poisson system is 
\begin{equation}
    \begin{cases}
        \displaystyle
        \frac{\partial \delta f_s}{\partial t}
    + \mathbf{v}\cdot\nabla_{\mathbf{x}}\delta f_s
    \displaystyle
    + \frac{q_s}{m_s}\delta\mathbf{E}\cdot
    \nabla_{\mathbf{v}} f_{s0}
    = C[\delta f_s], \\
    \nabla\cdot\delta\mathbf{E}
    =
    \displaystyle
    \frac{1}{\epsilon_0}
    \sum_s q_s\int \delta f_s\,d\mathbf{v}.
    \end{cases}
    \label{eq:vlasov-poisson}
\end{equation}
Here, $f_{s0}$ is the equilibrium distribution and $C[\delta f_s]$ represents a selected linear collision model.
$\delta f_s$ and $\delta\mathbf{E}$ denote the corresponding perturbations. 

\begin{figure*}
    \centering
    \includegraphics[width=\linewidth]{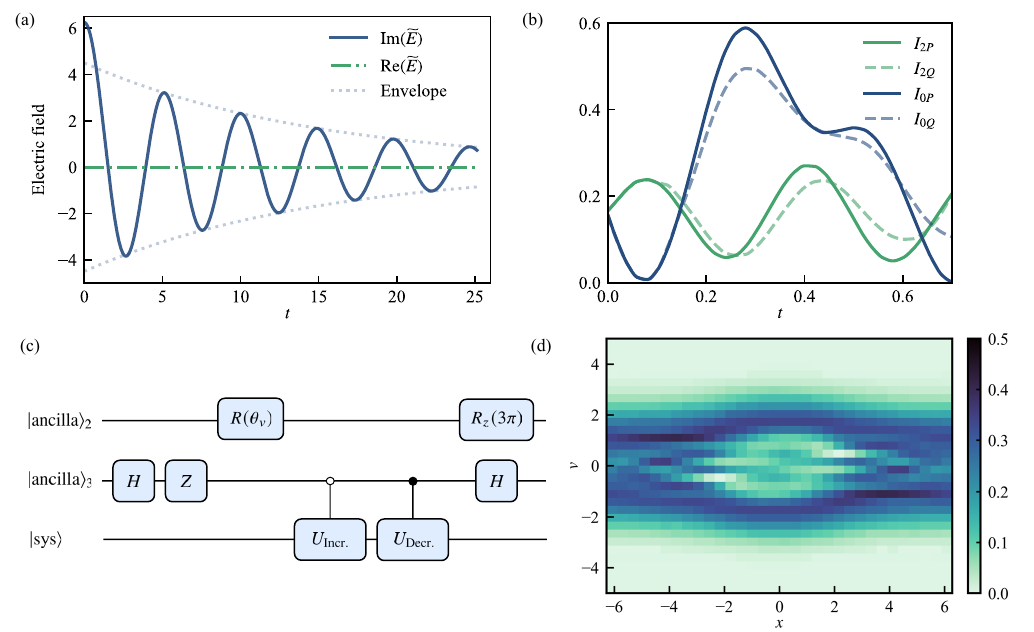}
    \caption{Results and circuit diagrams of studies of plasma transport.
            (a) The Landau damping result adapted from Fig.~5 in Ref.~\cite{engel2019}, where $\widetilde E$ denotes the Fourier component of the electric field for certain wave number, and the $y$-axis is scaled accordingly. 
            (b) Evolution of the normalized $l=0, 2$ mode populations in the pre-quantized $I_{lP}$ and second-quantized $I_{lQ}$ models, adapted from Fig.~5 of Ref.~\cite{may2025}.
            (c) The quantum circuit for the block encoding of the effective Hamiltonian of the 1D Vlasov equation adapted from Ref.~\cite{higuchi2025}, where $U_\mathrm{Incr.}$ and $U_\mathrm{Decr.}$ are unitary shift operators, $\mathrm{sys}$ denotes the system register encoding the spatial coordinate, $\mathrm{ancilla}$ denotes the auxiliary qubits.
            (d) Phase-space distribution of the 1D1V Vlasov-Maxwell system under two-stream instability conditions, adapted from Ref.~\cite{higuchi2025}.}
    \label{fig:vlasovResults}
\end{figure*}

For solving Eq.~\eqref{eq:vlasov-poisson}, most early quantum studies considered the collisionless limit $C[\delta f_s]=0$.
Engel et al.~\cite{engel2019} reformulated the original Vlasov-Poisson system into a unitary form 
by transforming the spatial coordinates to Fourier space and rescaling $f_s$. 
The velocity space was further discretized with finite difference, and the equation is solved via Hamiltonian simulation, in which the Hamiltonian is block-encoded through state-preparation unitaries.
Figure~\ref{fig:vlasovResults}(a) shows the Landau-damping result obtained in Ref.~\cite{engel2019}. 
The Landau damping provides a standard benchmark for linear collisionless kinetic models, in which electric-field oscillations decay through resonant wave-particle interactions.
Ameri et al.~\cite{ameri2023} instead adopted a Fourier-Hermite
representation, where the velocity dependence is expanded and truncated in a Hermite basis.
The system is again reformulated and solved by Hamiltonian simulation, while the Fourier-Hermite representation provides a more compact description. 
Toyoizumi et al.~\cite{toyoizumi2024hamiltonian} further studied the Hamiltonian-simulation approach using QSVT, with a detailed analysis of its error and query complexity. 
Paul et al.~\cite{paul2026efficient} compared Trotter-Suzuki and THRIFT Hamiltonian simulation methods for linearized Vlasov-Poisson dynamics, with particular emphasis on numerical accuracy and quantum circuit resources. 
Beyond these explicit quantum algorithms, Sugama~\cite{sugama2026schr} showed that both linear Vlasov-Poisson and linear gyrokinetic systems can be formulated as Schrödinger-type equations, providing a possible route toward quantum representations of more general collisionless kinetic models.

Parallel to the Vlasov-Poisson approach, linear collisionless kinetics can also be modeled using the Vlasov-Amp\`ere formulation. 
These models are often cast into matrix linear systems and solved via QLSA.
For example, Novikau et al.~\cite{novikau2024encoding} cast the linearized Vlasov-Amp\`ere system as a boundary-value linear problem and designed a compressed block encoding for the corresponding operator matrix. 
Goldfriend et al.~\cite{goldfriend2025implementation} subsequently provided an explicit QLSA implementation of this formulation, focusing on detailed circuit construction and resource overhead assessments.

Several studies have extended quantum kinetic simulations to collisional and dissipative regimes. 
Ameri et al.~\cite{ameri2023} incorporated a Lenard-Bernstein-type collision operator $C[g_k]=\nu\partial v \left((1/2)\partial v+v\right)g_k$ into the Fourier-Hermite formulation of the linearized Vlasov-Poisson equation, where $\nu$ is the collision frequency. 
Collisions introduce dissipative terms and make the evolution non-unitary.
Therefore the resulting linear system is solved using a quantum solver for ordinary differential equations rather than Hamiltonian simulation.
Amaro et al.~\cite{amaro2025variational} considered another dissipative kinetic problem based on the Fokker-Planck equation and applied variational quantum imaginary-time evolution to simulate the stochastic radiation-reaction cooling of an electron beam in a strong electromagnetic field.

Nonlinear Vlasov equations in Eq.~\eqref{eq:vlasov} are considerably more difficult to solve on quantum devices. 
Most existing studies are based on the Carleman linearization. 
Vaszary et al.~\cite{vaszary2024} applied this approach to a discretized Vlasov-Poisson equation with a Krook collision operator, and their analysis showed that the convergence condition imposes relatively strong requirements on the dissipation.
More recent studies have improved the formulation. 
Berntson et al.~\cite{berntson2026end} combined Fourier-Hermite expansion, a Lyapunov transformation, and Carleman embedding to obtain an end-to-end quantum algorithm for weakly nonlinear kinetic dynamics.
Christensen et al.~\cite{christensen2026improved} improved the convergence region toward more realistic collision frequencies. 
These approaches remain mainly applicable to weakly nonlinear and sufficiently dissipative regimes.

Alternative approaches either map nonlinear kinetic dynamics into linear evolutions within enlarged operator spaces or adopt quantum-classical hybrid frameworks. 
For instance, May and Qin~\cite{may2025} leveraged second quantization to formulate nonlinear Vlasov-Poisson dynamics as a linear system in an enlarged Fock space.
By representing interactions through creation and annihilation operators, this method avoids Carleman truncation at the expense of a truncated Fock space. 
Figure~\ref{fig:vlasovResults}(b) illustrates both the potentials and limitations of this second-quantization approach.
The quantized linear evolution captures the main mode population evolution of the nonlinear pre-quantized system, while visible discrepancies in long-time evolution highlight the need for sufficient quantum resolution to achieve quantitative agreement. 
Similarly, Tronci and Joseph~\cite{tronci2021koopman} explored Koopman-wavefunction formulations for the Vlasov-Maxwell system, casting classical kinetic dynamics into a Hilbert space with an explicit Hamiltonian structure. 
Alternatively, hybrid architectures isolate the system's nonlinearities by offloading high-dimensional linear transport to quantum algorithms while resolving nonlinear field interactions with classical processors.
This is demonstrated by Higuchi et al.~\cite{higuchi2025} in their QSVT-based Vlasov simulation coupled with a classical Maxwell solver.
The quantum circuit and result are illustrated in Figs.~\ref{fig:vlasovResults}(c) and (d), respectively.

These approaches address nonlinear kinetic dynamics through different representations and computational workflows. 
The Carleman-based algorithms introduce enlarged linear systems and remain primarily applicable to weakly nonlinear regimes under suitable convergence conditions, whereas second-quantized formulations require sufficient resolution in a truncated Fock space.
Hybrid approaches retain classical field solvers and therefore require quantum–classical data exchange. 
Their practical feasibility depends on different sources of overhead, including representation size, truncation error, and data exchange costs.

Overall, existing quantum algorithms for kinetic plasma dynamics have mainly focused on linearized Vlasov-Poisson systems or simplified nonlinear models. 
However, the major computational challenges in plasma transport arise from nonlinear gyrokinetic dynamics and high-dimensional phase space. 
Existing approaches generally do not simultaneously capture the global five-dimensional gyrokinetic dynamics, multispecies electromagnetic coupling, and collisional effects needed for realistic fusion-transport simulations.
Bridging this gap remains a challenge in extending quantum algorithm to more physically relevant plasma transport simulations.

\subsection{RF heating and current drive}
RF heating and current drive rely on the propagation of electromagnetic waves and the transfer of their energy and momentum to plasma particles. 
Rather than the complete RF heating and current-drive process in MCF plasmas, existing studies mainly consider simplified models among two categories. 
One is to simplify the plasma response and wave propagation, leading to linear models such as electromagnetic-wave propagation in cold magnetized plasmas. 
The other is to simplify the nonlinear wave dynamics while retaining the essential energy exchange among wave modes, leading to reduced models such as three-wave interactions.

\begin{figure*}
    \centering
    \includegraphics[width=\linewidth]{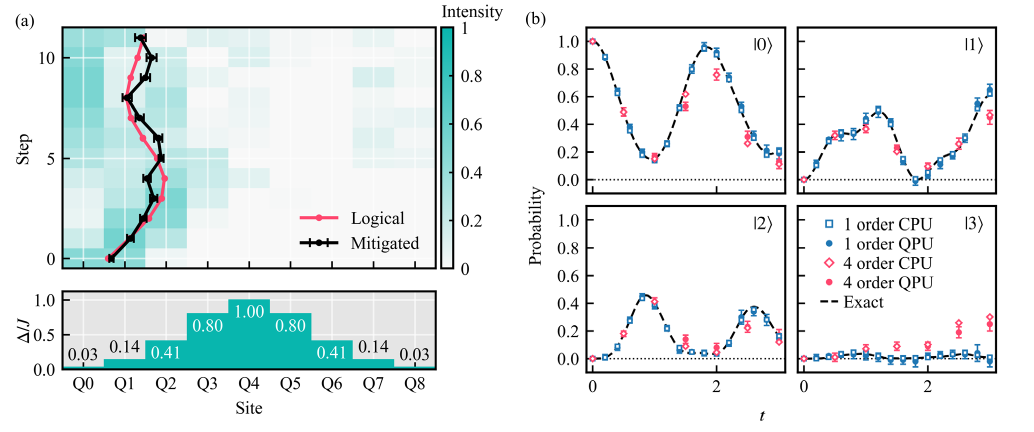}
    \caption{Hardware results for plasma wave simulations.
    (a) Error-mitigated propagation of an electromagnetic wave packet through an inhomogeneous overdense plasma with a Gaussian density profile, adapted from Fig.~4(c) of Ref.~\cite{sundar2026simulating}. 
    The upper panel shows the wave-packet intensity and center-of-mass trajectory, while the lower panel gives the normalized plasma-frequency profile. 
    (b) Time evolution of state occupation probabilities under combined three- and four-wave interactions simulated using first- and fourth-order product formulas at fixed gate depth, adapted from Fig.~3(a) of Ref.~\cite{shi2024simulating}. 
    The four panels represent states $|0\rangle$-$|3\rangle$ with different energy distributions among the waves.
    $|0\rangle$ is the initial state with only the pump excited, while higher-index states correspond to greater energy transfer to the other two waves. 
    }
    \label{fig:waveResults}
\end{figure*}

For linear wave propagation in cold magnetized plasmas, the propagation is described using the Maxwell equations with linearized plasma response
\begin{equation}
    \begin{cases}
        \displaystyle
        \frac{\partial \mathbf{B}}{\partial t}
        =
        -\nabla\times\mathbf{E}, \\
        \displaystyle
        \epsilon_0\frac{\partial \mathbf{E}}{\partial t}
        =
        \displaystyle
        \frac{1}{\mu_0}\nabla\times\mathbf{B}
        -
        \sum_s \mathbf{J}_s, \\
        \displaystyle
        \frac{\partial \mathbf{J}_s}{\partial t}
        =
        \epsilon_0\omega_{ps}^2\mathbf{E}
        +
        \Omega_{cs}\mathbf{J}_s\times\hat{\mathbf{b}}_0, 
    \end{cases}
    \label{eq:linearMaxwell}
\end{equation}
where $\mathbf{J}_s$ denotes the current density of species $s$, $\hat{\mathbf{b}}_0$ is the unit vector along the background magnetic field, and $ \omega_{ps}
    =
    \sqrt{n_{s0}q_s^2 / (\epsilon_0 m_s)}$ and 
    $\Omega_{cs}
    =
    q_s B_0/m_s$
are the corresponding plasma and cyclotron frequencies, respectively. 
Equation~\eqref{eq:linearMaxwell} can be reformulated as a Schr\"odinger-like equation, providing a natural starting point for quantum algorithms.
Novikau et al.~\cite{novikau2022quantum} considered a 1D cold-electron plasma immersed in a background magnetic field, and solved the resulting Schr\"odinger equation using QSP. 
Koukoutsis et al.~\cite{koukoutsis2023} extended the formulation to more general cold magnetized plasmas and applied a QLA to two-dimensional full-wave scattering.
They then developed a quantum-walk-inspired QLA with explicit circuit design and extended it to collisional dissipation using linear combinations of unitaries and post-selection~\cite{koukoutsis2025quantum}.
Hizanidis et al.~\cite{hizanidis2025spacetime} alternatively reformulated the original equations into a spinorial Dirac-type evolution equation with the Clifford geometric algebra of spacetime, which provides a natural representation for quantum simulation.
Meanwhile, Novikau et al.~\cite{novikau2023simulation} considered dissipative wave propagation as a non-Hermitian boundary-value problem and solved the resulting linear system using QSVT.
For hardware implementation, Sundar et al.~\cite{sundar2026simulating} adopted a different reduced model by mapping linear plasma-wave dynamics onto a local spin model, and demonstrated the resulting wave propagation on a superconducting quantum processor.
The error-mitigated wave packet propagation result is illustrated in Fig.~\ref{fig:waveResults}(a). 

In another major direction of nonlinear wave interactions, the three-wave interaction is one of the simplest models for describing nonlinear energy exchange among three resonantly coupled wave modes. 
The classical three-wave interaction is described by
\begin{equation}
    \begin{cases}
        \displaystyle
        \frac{\partial A_1}{\partial t}
        = g A_2 A_3, \\
        \displaystyle
        \frac{\partial A_2}{\partial t}
        = -g^* A_1 A_3^*, \\
        \displaystyle
        \frac{\partial A_3}{\partial t}
        = -g^* A_1 A_2^*,
    \end{cases}
    \label{eq:threeWaves}
\end{equation}
where $A_j$ denotes the complex amplitude of the $j$th wave and $g$ is the coupling coefficient.
By promoting the wave amplitudes to operators, Eq.~\eqref{eq:threeWaves} is then transformed to a cubic quantum Hamiltonian
$
    \hat{H}
    =
    \ii g \hat{A}_1^\dagger \hat{A}_2 \hat{A}_3
    -
    \ii g^* \hat{A}_1 \hat{A}_2^\dagger \hat{A}_3^\dagger .
$
Shi et al.~\cite{shi2021simulating} used conserved quantities to reduce this Hamiltonian to finite-dimensional invariant subspaces, and implemented the resulting Hamiltonian dynamics of a three-state example on a superconducting quantum processor.
May and Qin~\cite{may2024nonlinear} subsequently analyzed the quantum manifestation of classical three-wave instability~\cite{may2023quantum}, and then examined the long-time accuracy of the quantum representation. 
Shi et al.~\cite{shi2024simulating} extended the hardware implementation to three- and four-wave mixing processes and investigated practical circuit and error-mitigation strategies.
The results of first- and fourth-order product formulas displayed in Fig.~\ref{fig:waveResults}(b).

Moreover, Porter and Joseph~\cite{porter2025impact} employed the quantum sawtooth map, a periodically kicked quantum map that models chaotic particle diffusion in phase space, and implemented it on quantum hardware to investigate the transition from dynamical localization to chaotic diffusion.
Nonetheless, current studies address components of RF-related plasma dynamics, including wave propagation, reduced nonlinear wave interactions, and chaotic particle transport. 
Realistic coupling of wave propagation, particle dynamics, and energy deposition has not yet been addressed in current studies.

\subsection{Plasma equilibrium and MHD stability} 
Plasma equilibrium describes a stationary plasma configuration in which the macroscopic forces are balanced, and is commonly modeled using MHD equations. 
For example, in axisymmetric cases, the equilibrium MHD equations are reduced to the Grad-Shafranov equation~\cite{lodestro1994grad}, which is a nonlinear elliptic differential equation.
The equilibrium calculation therefore involves solving a nonlinear elliptic boundary-value problem for the magnetic configuration.
In 3D configurations, the equilibrium is instead commonly obtained through corresponding variational formulations. 

\begin{figure*}
    \centering
    \includegraphics[width=\linewidth]{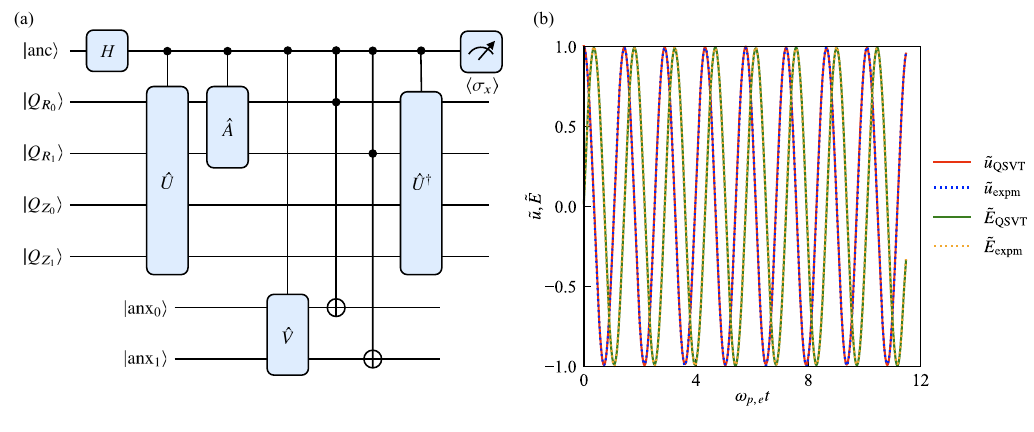}
    \caption{Results and circuit diagrams in plasma equilibrium studies.
    (a) Quantum circuit for evaluating the radial derivative contribution of the Grad-Shafranov operator, adapted from Fig.~3(a) in Ref.~\cite{mastorakis2026quantum}, where $Q_R$ and $Q_Z$ denote the radial and vertical coordinate registers, $\mathrm{anc}$ and $\mathrm{aux}$ denote the measurement and auxiliary qubits, respectively, and $\hat{U}$, $\hat{A}$, and $\hat{V}$ represent the parameterized ansatz, shift operator, and coefficient-encoding vector unitary, respectively.
    (b) Time evolution of the fluid velocity $\tilde{u}$ and electric field $\tilde{E}$ at $x=0$, adapted from Fig.~2 in Ref.~\cite{higuchi2025quantum}.
    The subscripts $\mathrm{QSVT}$ and $\mathrm{expm}$ denote results obtained using QSVT and classical matrix exponentiation, respectively.
    }
    \label{fig:equiResults}
\end{figure*}

Despite the importance of these problems, quantum algorithms for plasma-equilibrium calculations remain limited, with most existing studies focusing on the stability and dynamics of prescribed or simplified equilibria.
%
Ameri et al.~\cite{ameri2023quantum, ameri2024mhd, ameri2025quantum} 
linearized the ideal MHD equations around a prescribed equilibrium and formulated the stability analysis as an eigenvalue problem. 
They proposed a quantum framework based on block encoding and quantum phase estimation, with a potential polynomial speedup over classical algorithms.
Very recently, Mastorakis et al.~\cite{mastorakis2026quantum} solved the 2D Grad-Shafranov equation with VQA, where they developed two types of loss functions and compared their performance.
Figure~\ref{fig:equiResults}(a) shows the quantum circuit used in Ref.~\cite{mastorakis2026quantum} to evaluate the radial derivative contribution of the Grad-Shafranov operator.

Beyond equilibrium stability, several studies have investigated the dynamic evolution and instabilities of simplified MHD and macroscopic fluid models.
Vahala et al.~\cite{vahala2003lattice} carried out an early study of a highly reduced 1D, constant-density dissipative MHD model, for which they developed both lattice-Boltzmann and quantum lattice-gas representations. 
Designed for hybrid quantum-classical computing, this formulation was not demonstrated on actual quantum hardware at the time. 
It was later extended by Yepez et al.~\cite{yepez2005quantum} to handle arbitrary transport coefficients, including viscosity and resistivity.
More recently, Higuchi et al.~\cite{higuchi2025quantum} investigated nonlinear instability in a reduced electromagnetic-fluid model. 
They considered an incompressible, pressureless fluid coupled to Maxwell's equations, linearized the nonlinear dynamics using the Koopman-von Neumann formalism, and simulated the resulting evolution using QSVT-based Hamiltonian simulation. 
Figure~\ref{fig:equiResults}(b) compares the resulting fluid-velocity and electric-field evolution with classical matrix-exponentiation results given by Higuchi et al.~\cite{higuchi2025quantum}.

In short, quantum applications in plasma equilibrium and stability have received less attention compared to other subdomains.
Existing studies are limited to simplified equilibrium models, linear stability analysis, and reduced MHD dynamics.
Moreover, the readout cost depends on whether the task requires only selected eigenvalues or observables, or the reconstruction of spatially resolved fields.
It remians challenging to extend these approaches to more realistic plasma models.

\subsection{Other applications and open problems}
\begin{figure*}
    \centering
    \includegraphics[width=\linewidth]{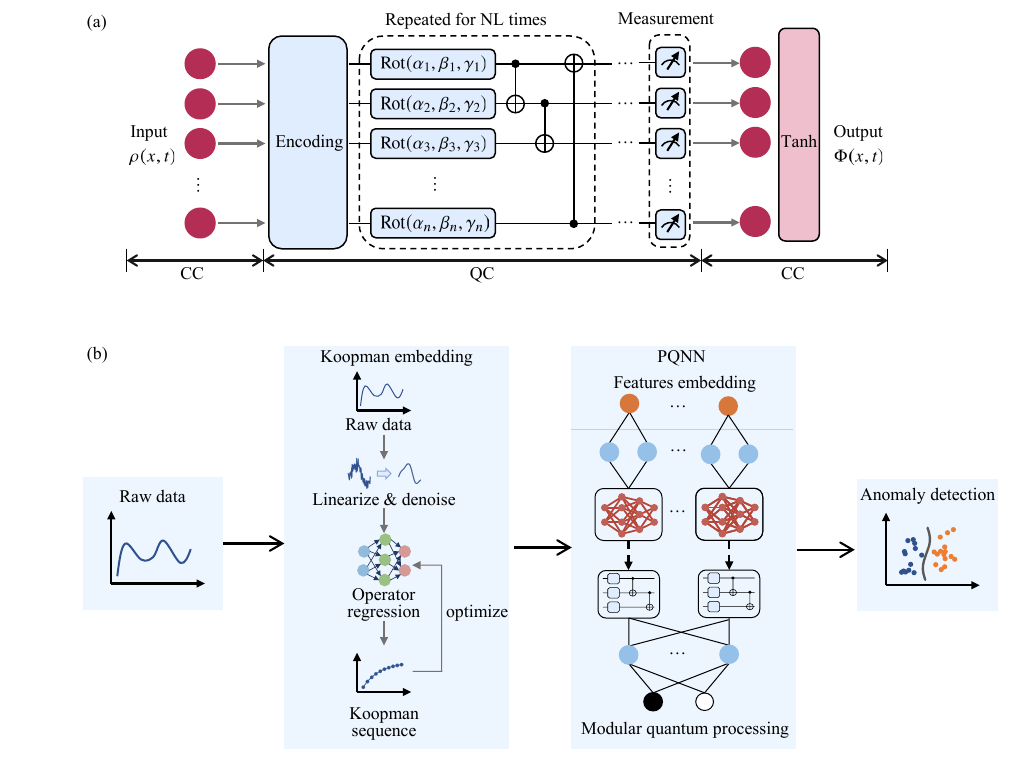}
    \caption{Diagrams of plasma-related quantum machine learning studies.
    (a) The classical-quantum-classical diagram to solve the Poisson equation adapted from Fig.~1 in Ref.~\cite{hegde2025hybrid}, where the input is the charge density $\rho(x, t)$, the output is the electric potential $\Phi(x, t)$, CC denotes classical computing, and QC denotes quantum computing.
    (b) The architecture of the Koopman-PQNN framework adapted from Fig.~2 in Ref.~\cite{wang2026validating}, where Koopman features extracted from diagnostic signals are processed by parallel quantum circuits for anomaly detection.}
    \label{fig:otherResults}
\end{figure*}

Beyond direct numerical solvers for kinetic, wave, and MHD dynamics, several studies have explored ancillary applications of quantum computing in plasma research. 
Rather than replacing complete plasma solvers, these approaches typically embed quantum neural networks as trainable components within predominantly classical workflows. 
For instance, Hegde et al.~\cite{hegde2025hybrid} presented a quantum-assisted PIC scheme that uses a quantum neural network to solve the Poisson equation for the electrostatic potential given grid-based charge densities, while retaining classical particle pushing.
Their quantum-classical architecture used to approximate the electrostatic potential is illustrated in Fig.~\ref{fig:otherResults}(a).
Similarly, Wang et al.~\cite{wang2026validating} combined Koopman embedding with a parallel quantum neural network (PQNN) for anomaly detection in tokamak diagnostic signals, as illustrated in Fig.~\ref{fig:otherResults}(b).

Despite this algorithmic diversity, existing research primarily considered simplified geometries and reduced models.
High-fidelity scenarios featuring strong nonlinearities, intricate boundaries, and cross-scale dynamics have not been demonstrated in studies reviewed here. 
We categorize these open problems into three primary domains. 
The first addresses high-fidelity multiscale modeling, encompassing gyrokinetic turbulence, edge and SOL dynamics, plasma-wall interactions, and coupled core-edge-wall transport. 
The second focuses on rare-event analysis and uncertainty quantification, such as runaway-electron events, wall-impact probabilities, transport under uncertain plasma parameters.
The third targets integrated facility operation, including the coupling of plasma models with diagnostics, prediction, optimization, and ultimately real-time control.
Overall, current plasma quantum computing is largely restricted to simplified models and isolated subroutines, highlighting a gap between existing proof-of-concept demonstrations and the rigorous demands of fusion energy applications.

\section{Challenges, opportunities, and roadmap}
\label{sec:roadmaps}
\subsection{Algorithmic challenges in quantum computing for plasma simulation}
Existing quantum studies of plasma physics have largely been restricted to linearized kinetic equations, reduced wave interaction models, linear stability analysis, and simplified fluid dynamics. 
This is partly because incorporating more realistic physics makes it harder to preserve or exploit structures that enable efficient quantum algorithms, such as linearity, sparsity, and efficient operator representations.
The present gap between current quantum studies and practical fusion applications therefore arises from several closely related physical and computational barriers.

A first challenge concerns the nonlinear and non-unitary structures of realistic plasma models. 
Taking kinetic equations as an example, away from equilibrium, the distribution function evolves self-consistently with the electromagnetic fields. 
Since the fields themselves depend on the particle distribution, terms such as the electric-field acceleration in the Vlasov equation introduce quadratic nonlinearities into the coupled system. 
Collisions further introduce dissipative and therefore non-unitary dynamics.
Moreover, collision operators used in realistic kinetic simulations are themselves generally nonlinear, adding another layer of difficulty for quantum implementation. 
Nonlinearity becomes even more prominent in macroscopic models such as MHD and in equilibrium calculations based on the Grad-Shafranov equation. 
These features make realistic plasma models substantially more difficult to accommodate within existing quantum algorithms, which are most mature for linear and unitary dynamics.

Another major difficulty arises from initial-state preparation and final-state measurement.
The exponential representation capacity of quantum states does not directly translate into an exponential computational speedup.
Preparing an initial quantum state that encodes the problem data can incur substantial computational costs~\cite{mottonen2004transformation}, while extracting useful information from the final state generally requires repeated measurements~\cite{preskill2018quantum}.
These input and output costs can offset the advantage gained during quantum evolution.

Challenges also arise from the coupling of models across different physical scales and from realistic geometries and boundary conditions. 
For example, edge and SOL simulations often couple fluid transport models with Monte-Carlo descriptions of neutral particles, while kinetic or PIC models may be required closer to plasma-facing materials.
Similarly, simulations of RF heating and current drive require coupling full-wave electromagnetic solvers with kinetic models for particle distributions. 
Efficient quantum implementations of many individual plasma models remain an open technical challenge, and integrating multiscale and multiphysics models therefore requires addressing additional issues.
Realistic boundaries introduce an additional obstacle.
Many existing quantum algorithms for linear differential equations rely on Fourier or other structured representations, which are most natural for simple or periodic domains. 
Finite-difference discretizations can treat more general boundaries, but they introduce larger and less structured operators together with additional encoding overhead.
Therefore, applying quantum computing to realistic fusion devices will require not only more complete quantum solvers, but also efficient treatments of model coupling, complex geometries, interfaces, and boundary conditions.

Note that, in addition to the computational challenges discussed in Sec.~\ref{sec:introChallenges}, quantum computing requires specific mathematical properties, thereby imposing additional requirements on problem formulation and implementation. 
Accordingly, directly replacing general-purpose plasma solvers with quantum algorithms does not by itself establish a computational benefit. 
Achieving a practical quantum advantage requires identifying plasma problems with structures compatible with quantum algorithms and accounting for the costs of reformulation, state preparation, and measurement.
Such assessments should also account for errors from discretization, model reduction or truncation, quantum algorithm approximations, hardware noise, and measurement, evaluated in terms of their impact on the target physical observables.

\subsection{Promising problems for quantum advantage} 
We expect plasma problems with characteristics compatible with quantum computing to be more suitable candidates for practical quantum advantage. 
First, the problem involves a large-scale or high-dimensional state space that is computationally expensive to store or process using classical methods.
Second, due to the linear and unitary nature of quantum devices, the governing equations are preferably linear or weakly nonlinear, or amenable to efficient linearization, while the resulting operators should remain sparse or otherwise structured for efficient quantum representation and implementation.
Finally, the costs of initial-state preparation and final-state measurement should remain sufficiently small that they do not dominate the overall computational complexity or eliminate the potential quantum advantage.

The kinetic models provide a useful illustration of these criteria as introduced in Sec.~\ref{sec:currentResearch}. 
They describe particle distribution functions in high-dimensional phase space, which creates substantial storage and computational costs for classical methods, while some components of their dynamics retain linear or otherwise structured forms. 
Although realistic kinetic models can still involve strong nonlinear field coupling and collisions, certain problems permit these effects to be separated, approximated, or treated through hybrid approaches. 
Meanwhile, other macroscopic models such as MHD involve additional nonlinear coupling of different physical quantities, complex geometries, and boundary conditions.
Nonetheless, these criteria can also be met by reformulations of other plasma problems. 

Next, we discuss four possible routes toward quantum advantage in plasma simulation, including accelerating computational  components, reformulating plasma problems, seeking alternative quantum representations, and pursuing dedicated quantum implementations.
Rather than directly replacing a complete plasma solver with a quantum algorithm, these approaches seek either to identify plasma problems whose original structures are already compatible with quantum computation, to reformulate difficult dynamics into more quantum-friendly representations, or to accelerate computational subroutines within otherwise classical workflows. 

First, a relatively direct route is to use quantum algorithms to accelerate Monte-Carlo components within existing plasma simulation workflows. 
Monte-Carlo procedures already appear in several important plasma computation codes~\cite{wiesen2015new, frerichs2020detachment, romazanov2020first}. 
For example, in kinetic simulations, the collisional term $C[f_s]$ in Eq.~\eqref{eq:vlasov} can be treated through stochastic collision sampling. 
In edge and SOL modeling, neutral particle transport is also commonly solved by Monte-Carlo methods and coupled to fluid plasma equations. 
These problems often involve sampling trajectories or interactions in high-dimensional phase space, while only a limited number of statistical quantities, such as collision rates, transport fluxes, or loss probabilities, are ultimately required.
The combination of high-dimensional sampling and limited statistical output makes such problems attractive for quantum Monte-Carlo methods, which can in principle provide a quadratic speedup in estimating probabilities and expectation values. 
However, the end-to-end advantage depends critically on whether the underlying Monte-Carlo dynamics can be efficiently represented and implemented on quantum hardware, as well as on the costs of state preparation and measurement.

Second, certain plasma problems can be reformulated in quantum-compatible forms such as stochastic differential equations (SDEs).
A major obstacle to quantum simulation of plasma dynamics is their strong nonlinearity, which is difficult to handle efficiently with current linearization, truncation, or hybrid approaches.
For the SDE formulation, the evolution or target observables can admit linear or structured representations, which is favorable for quantum computing.
Recent studies have also begun to address nonlinear SDEs~\cite{miyamoto2026quantum, li2026efficient, bravyi2026quantum}. 
Such formulations therefore offer a promising way to treat certain nonlinear plasma processes, potentially reducing reliance on conventional linearization or repeated quantum-classical feedback.

Third, a more fundamental route is to seek alternative quantum representations of classical plasma dynamics rather than directly solving the original governing equations.
Since quantum computing was originally motivated by the simulation of quantum systems, one possibility, as suggested by Joseph et al.~\cite{joseph2023}, is to represent classical dynamics using an interacting quantum system whose evolution reproduces the desired macroscopic behavior. 
A representative example is the quantum formulation of classical three-wave interactions, where classical wave amplitudes are promoted to quantum operators and the resulting quantum dynamics can reproduce characteristic nonlinear energy exchange over finite time scales~\cite{shi2021simulating, shi2024simulating}. 
This approach is conceptually related to linearization or embedding, but instead exploits the intrinsic structure and interactions of the quantum system itself.
While such representations may not be broadly applicable~\cite{may2024nonlinear}, suitable plasma problems still admit effective quantum representations and provide an alternative route to nonlinear dynamics.

Fourth, plasma dynamics with suitable mathematical structures can be implemented directly using analog quantum systems whose native interactions reproduce the desired evolution.
For example, continuous-variable quantum simulators have been proposed to implement Schr\"odingerized equations directly through physical Hamiltonians, including formulations for Fokker-Planck and Maxwell equations, while pseudospin-boson systems have been proposed to reproduce lattice-kinetic transport dynamics. 
Nonlinear quantum systems such as Bose-Einstein condensates also naturally realize Gross-Pitaevskii-type dynamics. 
These examples suggest that, if a plasma model can be transformed into a form compatible with the native dynamics of an experimentally accessible quantum system, analog quantum platforms can provide an alternative to gate-based implementations.

In summary, practical quantum advantage in plasma physics remains an open direction.
Achieving an end-to-end quantum advantage will ultimately depend on the structure of the specific plasma problem, as well as on whether the costs can remain below those of the best available classical approaches.

\subsection{Hardware demonstrations and resource requirements}
As a complement to the algorithmic discussion, we briefly introduce several major quantum computing hardware platforms and their potential relevance to plasma simulation. 
Superconducting qubits, trapped ions, and neutral atoms are currently among the most prominent platforms for quantum computation~\cite{vukvsic2026comparative}.

Superconducting quantum processors use circuits containing Josephson junctions as artificial atoms and operate at cryogenic temperatures to suppress thermal noise. 
Current state-of-the-art superconducting processors have demonstrated more than one hundred high-fidelity physical qubits~\cite{gao2025zuchongzhi}, while a recent distance-7 surface-code experiment used 101 physical qubits to realize one logical qubit~\cite{google2025quantum}.
Gate operation times typically range from a few nanoseconds to hundreds of nanoseconds, with over $99.9\%$ two-qubit gate fidelities~\cite{bland2025millisecond,li2024realization}. 
Nevertheless, residual gate errors limit practical circuit depths to tens of cycles~\cite{gao2025zuchongzhi}.
These characteristics make superconducting processors particularly suitable for relatively shallow gate-based, variational, and hybrid algorithms, while deep fault-tolerant algorithms require further scaling of logical qubits.

Trapped-ion quantum processors use charged ions confined by electromagnetic fields as qubits, with quantum operations implemented using laser or microwave fields.
Current trapped-ion processors have approached one hundred physical qubits with nearly all-to-all connectivity~\cite{ransford2026helios}, while circuits involving up to 12 logical qubits have been demonstrated~\cite{paetznick2026improved}.
Two-qubit gate fidelities have exceeded \(99.9\%\) in current processors, while the two-qubit gate times of approximate $70~\mu\mathrm{s}$ are longer than those in superconducting hardware.
Trapped-ion processors can offer exceptionally long coherence times, exceeding one hour in single-ion quantum memories~\cite{wang2021single}. 
Their long coherence, high fidelity, and flexible connectivity therefore make trapped-ion processors attractive for high-accuracy and relatively deep circuits, particularly those involving nonlocal interactions and quantum error correction.

Neutral-atom quantum processors use individually trapped atoms as qubits and exploit controllable Rydberg interactions to implement quantum operations.
Neutral-atom arrays have demonstrated strong scalability, with up to 280 physical qubits and 48 logical qubits on programmable processors~\cite{bluvstein2024logical}, while recent  experiments demonstrated surface-code error correction on 448 atoms~\cite{bluvstein2026fault}.
Neutral-atom qubits also offer long coherence times of up to $12.6$s~\cite{manetsch2025tweezer}, together with two-qubit gate fidelities over \(99.5\%\)~\cite{evered2023high,finkelstein2024universal}. 
The main advantages include scalability, reconfigurable connectivity, and native programmable interactions, while atom loss, gate fidelity, and the overhead associated with atom transport and large-scale control are important challenges of practical neutral-atom devices. 
These properties make neutral-atom processors particularly attractive for large-scale quantum simulation, structured Hamiltonian evolution, and algorithms that benefit from flexible connectivity and parallel interactions.

In summary, these platforms offer complementary trade-offs in speed, fidelity, coherence, connectivity, and scalability. 
Current NISQ devices can support proof-of-principle simulations of reduced plasma models, whereas deep quantum algorithms and realistic fusion applications will require further advances in fault-tolerant hardware.

\subsection{Roadmap} 
\begin{figure*}
    \centering
    \includegraphics[width=\linewidth]{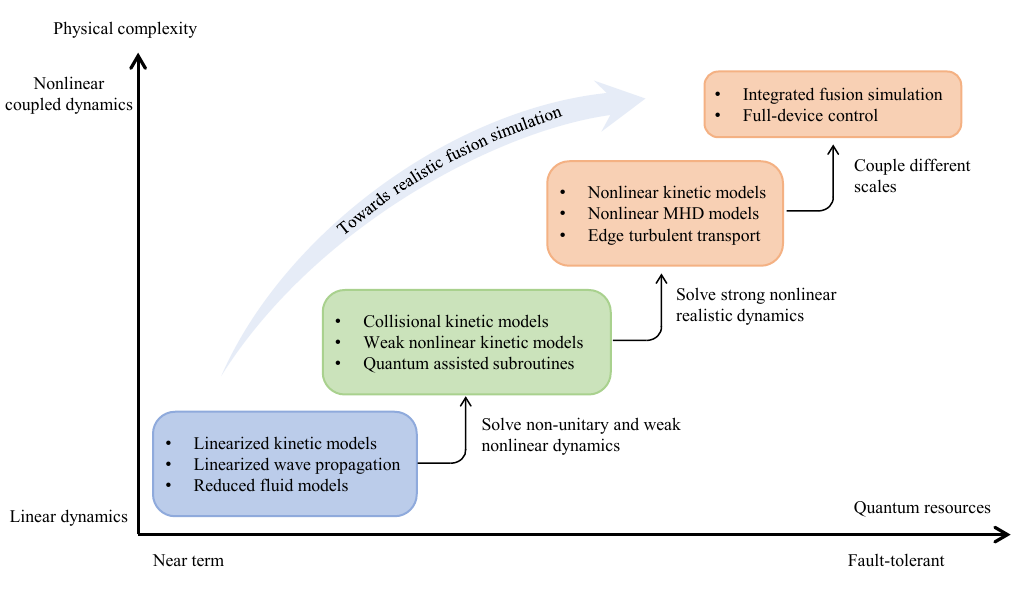}
    \caption{Roadmap toward practical quantum plasma simulation for fusion applications.
    The horizontal and vertical axes represent increasing quantum-computing resources and physical complexity, respectively. 
    Blue, green, and orange regions indicate representative problems for the near-term, intermediate, and long-term stages. 
    The arrows highlight the main challenges associated with progressing toward realistic fusion simulations.}
    \label{fig:roadmap}
\end{figure*}

Based on the algorithmic opportunities and hardware capabilities discussed above, we provide a possible roadmap toward quantum computing for plasma simulation.
The roadmap is divided into three stages, according to the scale and complexity of the plasma problems that can be addressed and the maturity of quantum hardware.
The overall roadmap diagram is illustrated in Fig.~\ref{fig:roadmap}.

In the near term (e.g., in five years), we expect further development of quantum algorithms for reduced linear models, SDE-based formulations, and simple quantum-classical hybrid methods.
This stage should also include small-scale hardware demonstrations to validate quantum representations, algorithms, and measurement strategies for physically relevant plasma problems.

In the intermediate term, quantum algorithms become capable of treating larger-scale linear problems, non-unitary dynamics, more complex plasma-wave interactions, and increasingly efficient hybrid simulations.
Hardware demonstrations should also move from simplified models toward more realistic applications.
At this stage, quantum subroutines begin to complement classical plasma codes in some computationally intensive tasks, such as statistical estimation, optimization, and reduced kinetic calculations.

In the long term, the development of fault-tolerant quantum hardware can enable end-to-end algorithms for strongly nonlinear dynamics, multiphysics coupling, and realistic geometries.
For operational fusion devices, quantum computing could then contribute not only to high-fidelity predictive simulations, but also to fast equilibrium and stability analysis, scenario optimization, uncertainty assessment, and eventually real-time plasma prediction and control.
Rather than replacing all classical computation, such applications will likely rely on integrated quantum-classical workflows that combine quantum acceleration with established fusion modeling, diagnostics, and control systems.

\section{Conclusions}
\label{sec:Conclusions}
We examined quantum computing for plasma physics from the perspective of computational problems in MCF. 
Existing studies have explored quantum algorithms for kinetic equations, plasma-wave dynamics, stability analysis, reduced fluid models, and related tasks. 
Among these, kinetic formulations have received considerable attention in recent quantum studies because they combine high-dimensional phase spaces with structured dynamics in certain cases. 
However, most demonstrations have not yet incorporated realistic nonlinearities, dissipation, self-consistent field coupling, complex geometries, and cross-scale physics.

Conditional end-to-end theoretical speedups have been reported for selected reduced plasma models. 
However, these results should be distinguished from experimentally demonstrated practical advantages in realistic fusion-simulation workflows.
Further progress therefore requires identifying plasma problems and computational approaches that are particularly compatible with quantum computing, together with continued developments in algorithms, representations, and hardware implementations. 
Promising directions include high-dimensional kinetic calculations, quantum Monte-Carlo methods, stochastic formulations, and alternative quantum representations of classical plasma dynamics. 
These developments could support the extension of quantum-plasma studies from proof-of-concept demonstrations and reduced models toward realistic fusion scenarios and tightly coupled quantum-classical hybrid workflows. 
Ultimately, practical quantum advantage in plasma physics should be assessed at the level of the complete computational workflow and against the best available classical approaches for the same physical outputs and target accuracy.

\section*{Acknowledgments}
This work has been supported in part by the National Natural Science Foundation of China (grant Nos.~12525201, 12432010, 12588201, 12532013, and 12588301). 

\bibliographystyle{elsarticle-num-names} 
\bibliography{ref}

\end{document}